\documentclass[%
reprint,
 amsmath,amssymb,
 aps,
]{revtex4-2}

\usepackage{graphicx}
\usepackage{dcolumn}
\usepackage{bm}

\begin{document}


\title{Bohmian Trajectories in a Bistable Potential Well }

\author{O. F. de Alcantara Bonfim}
\affiliation{%
Department of Physics, University of Portland, Oregon 97203 
}%


\begin{abstract}
We analyze the dynamics of a quantum particle in a one-dimensional bistable
potential within the framework of Bohm's quantum mechanics.
We give arguments that evidence the fallacy of certain claims found in the 
literature dealing with the impossibility of chaotic behavior of Bohmian 
trajectories in one-dimensional systems.
We find that an appropriate choice for the initial position and wave packet
causes the particle to undergo periodic, quasiperiodic, or chaotic motion.
The transitions between these regimes occur in a continuos
fashion.  

\end{abstract}

\keywords{Chaos, Bohm's quantum mechanics, bistable potential, Lyapunov exponents}
\maketitle


\section {\label{sec:1.} Introduction}

Chaotic behavior in classical systems is characterized by a sensitive
dependence on initial conditions. This unique feature leads to an exponential
divergence of trajectories and subsequent unpredictability in the behavior of
the system. This characterization of chaos cannot, however, be extended to
the quantum realm using conventional quantum mechanics due to the 
non-existence of the concept of trajectory in its formulation. 
Nonetheless, there is a great deal of interest in the behavior of
quantum systems whose classical counterparts are chaotic.  In particular,
one looks for unusual features which might be a leftover from the
chaotic behavior of the classical system.  Several fingerprints have
been proposed to identify the manifestation of chaos in the quantum
domain \cite{books}. The term quantum chaos refers usually to the study 
of quantum systems that have chaotic classical counterparts.

In the early 50's, David Bohm put forth a theory for the microscopic realm,
in which there appears a concept for the particle's trajectory.  The 
theory is based on Schroedinger equation and was constructed in such a way
so as to reproduce all of the results of conventional quantum mechanics.
Such a feat is achieved after a judicious averaging over the 
initial positions of all possible trajectories.

The so-called quantum potential that appears in the diferential equation 
for the particle trajectory is usually nonlinear as well as time-dependent, 
thus deeming analytic solutions very hard to come by.  In cases like that, 
one would expect to encounter a wide class of solutions, ranging from 
periodic to chaotic solutions. 
In fact, we can find several examples of such behaviors in the literature
\cite{our paper and others}.  Note that the nonlinearity of the equations for the 
Bohmian trajectories is a general feature, regardles the spatial dimension.  
There are numerical evidence of chaotic as well as regular behavior in one- 
and two-dimensions \cite{specific papers on chaos in one and two d}.

Note that one could find chaotic behavior in Bohmian trajectories in 
systems whose classical counterparts are regular.  That is, chaoticity is
an intrinsic feature of the quantum trajectory, and does not depend 
on whether or not the classical counterpart is chaotic.

Another way to investigate quantum 
chaos is to adopt the view of some authors \cite{Bohm93,Hol93} who 
have suggested that the quantum trajectories present 
in Bohm's formulation of quantum mechanics 
may be expected to be sensitive to initial conditions.  In fact, the existence 
of well-defined particle trajectories in Bohm's theory makes it a natural 
choice for the investigation of chaos in quantum systems. In this way the 
classical definition of chaos is retained, and all the tools used to 
identify chaos in classical systems are also readily available for the study 
of quantum chaos.

In the recent years, there has been a great deal of interest on the
existence of Bohmian trajectories showing chaos, as well as on the 
conditions needed for such behavior
\cite{Sal01,Sch95,Par95,Fai95,Pol96,Sen96,Iac96,Dew96,Fri96,Sch97,Kon98,Bon98a,Bon98b,Bon00,Sal03}.
The application of Bohm's mechanics to the one-dimensional quantum
delta-kicked rotator \cite{Sch95,Dew96} led to a surprising result: 
albeit the classical
counterpart of the delta-kicked rotator is chaotic, the quantum version shows
no chaotic behavior. Quantum effects seem to suppress the otherwise
chaotic behavior in the system. 
Dewdney and Malik \cite{Dew96} have shown, however, that the
effect of repeated measurements leads the system to behave chaotically. 

A more interesting picture emerges when chaos arises from the intrinsic
dynamics of the system and not as a result of some external measurements 
or as a response to external random forces.  In that regard, 
Parmenter and Valentine \cite{Par95} have shown that chaotic behavior is
present in the two-dimensional anisotropic harmonic oscillator which,
incidentally, is
not chaotic in the classical version. They also 
use general arguments based on the Poincar{\'e}-Bendixson theorem to rule out
chaos in one-dimensional systems.
In the present work we would like to address this problem by first 
questioning the argument of the absence of chaos in one-dimension, and then
by presenting  a one-dimensional 
system that clearly displays chaotic behavior.

In a previous work, we investigated the dynamics of
a particle in a square double-well potential and observed chaotic motion for a
suitable choice of the initial wave packet \cite{Bon98b}.
However, one might argue
that the seemly chaotic behavior observed is caused by the fact that the
potential is discontinuous. Here we show that chaotic behavior is also
present in the case of a continuos one-dimensional potential, thus the
discontinuity in the potential is not an essential ingredient for chaos
in Bohmian trajectories.

\begin{figure}
\begin{center}
\includegraphics[height=10cm]{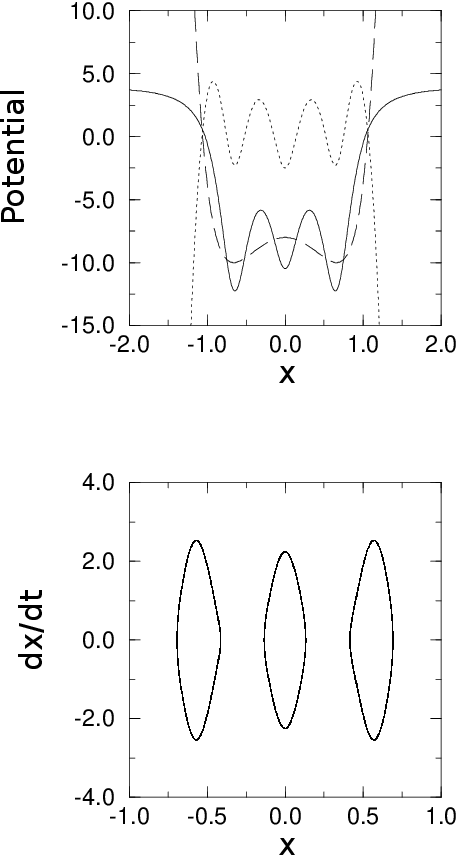}       
\caption{
(a) Potential as a function of the position, for $\xi= 4$.
The dashed line is the  bistable potential with maximum at the origin and 
minima at $x= \pm 0.66$. The system of units used is such that $\hbar=M=1$,
and the distances are measured in units of $\beta^{-1}$.
The dotted line is the quantum potential at $t=0$, corresponding to
the initial wave packet $\psi(x)= iu_{0}(x) + 10\,u_{3}(x)$.
The solid line is the effective potential $V_{eff}= V + Q$.
(b) Particle motion under the action of the effective potential.
The phase-space plots are for different initial conditions: left ($x_{0}= -0.5$),
center ($x_{0}= 0$), and right ($x_{0}= 0.5$). 
}                                            
\label{fig:fig1}                              
\end{center}                                 
\end{figure}

\section{\label{sec:2.} The Model}
Let us consider the one-dimensional motion of a quantum particle of
mass $M$ under the action of the following bistable potential 

\begin{equation}
V(x)= \frac{{\hbar}^2{{\beta}^2}}{2M}{\xi}\,[\frac{{\xi}}{8}(\cosh(4x) -1)
           - (n+1)\cosh(2x)],
\label{eq:potential}
\end{equation}
where $\beta$ and $\xi$ are arbitrary parameters
and $n$ is an integer\cite{Raz80}. The form 
of this potential allows us to analytically obtain the first $n+1$ 
eigenvalues and eigenfunctions. For $n= 3$ the eigenvalues are given 
by $E_n= ({\hbar^2}{\beta}^2/2M)\epsilon_n $, with ${\epsilon}_{n}$
given by

\begin{equation}
         {\epsilon}_{0}= -5 - \xi - {\alpha}_{-}, 
\label{eq:e0}
\end{equation}
\begin{equation}
         {\epsilon}_{1}= -5 + \xi - {\alpha}_{+}, 
\label{eq:e1}
\end{equation}
\begin{equation}
         {\epsilon}_{2}= -5 - \xi + {\alpha}_{-}, 
\label{eq:e2}
\end{equation}
and
\begin{equation}
         {\epsilon}_{3}= -5 + \xi + {\alpha}_{+}, 
\label{eq:e3}
\end{equation}
where ${\alpha}_{\pm}= 2(4 \pm 2\xi + {\xi}^2)^{\frac{1}{2}}$. The 
corresponding eigenfunctions are
$u_{n}(x)= \exp[-(\xi/4) \cosh(2x)]{\phi}_{n}(x)$, with ${\phi}_{n}(x)$
given by  

\begin{equation}
         {\phi}_{0}(x)= 3{\xi}\cosh(x) + (4 - \xi + {\alpha}_{-})\cosh(3x), 
\label{eq:f0}
\end{equation}
\begin{equation}
         {\phi}_{1}(x)= 3{\xi}\sinh(x) + (4 + \xi + {\alpha}_{+})\sinh(3x), 
\label{eq:f1}
\end{equation}
\begin{equation}
         {\phi}_{2}(x)= 3{\xi}\cosh(x) + (4 - \xi - {\alpha}_{-})\cosh(3x), 
\label{eq:f2}
\end{equation}
and
\begin{equation}
         {\phi}_{3}(x)= 3{\xi}\sinh(x) + (4 + \xi - {\alpha}_{+})\sinh(3x). 
\label{eq:f3}
\end{equation}

\begin{figure}
\begin{center}
\includegraphics[height=16cm]{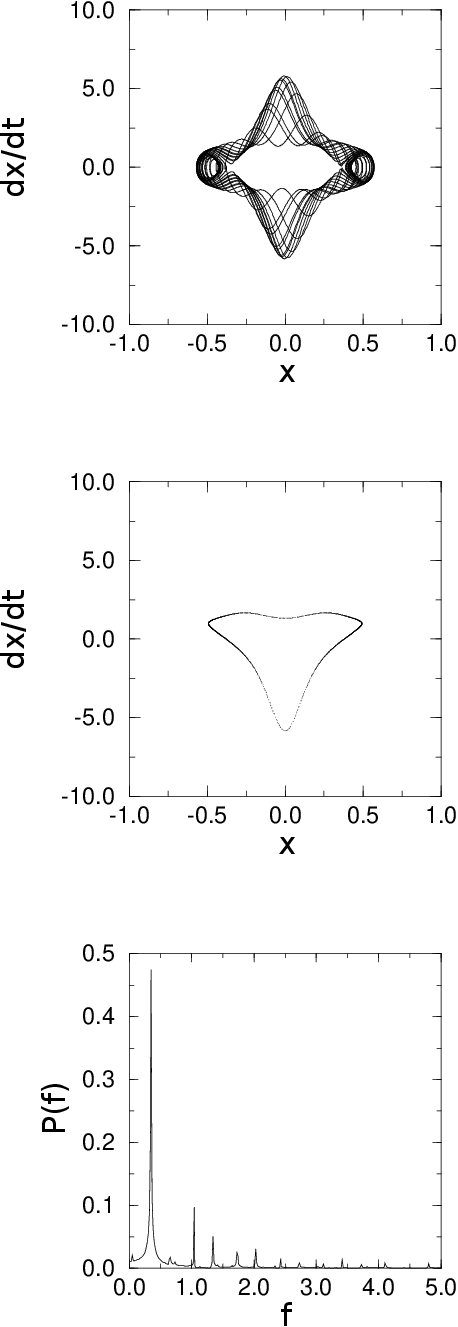}     
\caption
{
(a) Phase-space portrait of the motion of a quantum particle in a bistable
potential with initial wave packet 
$\psi(x,0)= iu_{0}(x) + u_{1}(x) + 10\,u_{3}(x)$ and initial position
$x_{0}= 0$. The distances are measured in units of  
$\beta^{-1}$ and $\hbar=M=1$.
(b) Stroboscopic view of the phase portrait. Positions
and velocities are shown at times multiple of $T=2\pi/{\omega}$,
with $\omega= (E_{3} - E_{1})/{\hbar}$.
(c) Power spectrum as a function of frequency $f={\omega}/{2\pi}$
obtained from the time series of $x(t)$. 
}                                            
\label{fig:fig2}                              
\end{center}                                 
\end{figure}

One should note that bistable potentials have been used as models for
diatomic ionic molecules,
where the two minima represent the centers of force in which the motion 
of a particle is taking place.  In addition, the solutions to the Schr\"odinger 
equation with this kind of potential has been used in the study of 
classical diffusion \cite{Kam77}.

\begin{figure}
\begin{center}
\includegraphics[height=16cm]{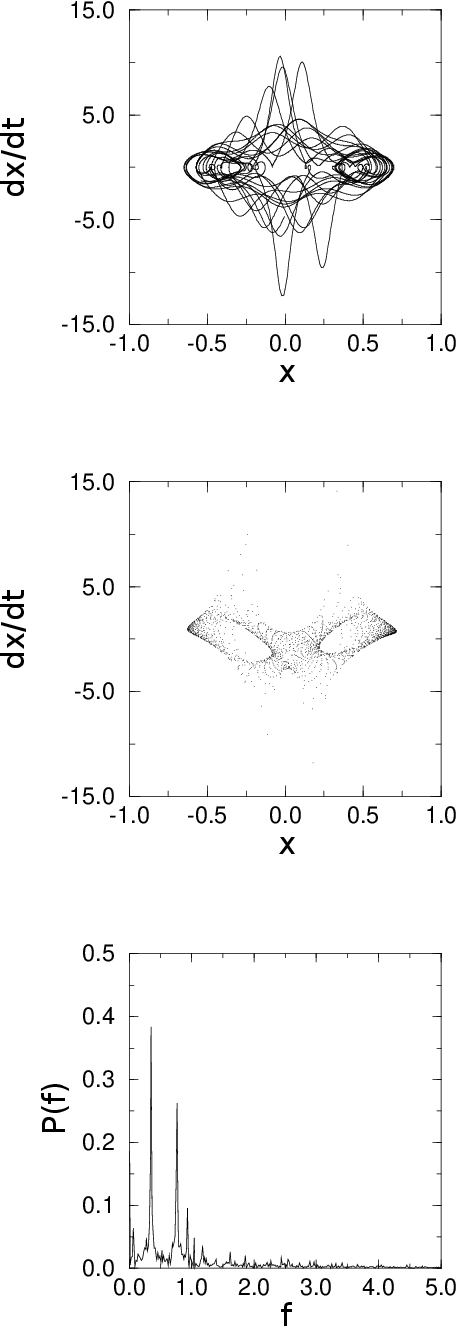}       
\caption
{
(a) Phase-space plot for a quantum particle in a bistable potential.
The particle position at $t=0$ is $x_{0}= 0$, and the initial wave packet
is $\psi(x,0)= iu_{0}(x) + u_{1}(x) + 4\,u_{2}(x) + 10\,u_{3}(x)$. 
In the system of units used, 
$\hbar=M=1$, and the lengths are measured in units of $\beta^{-1}$.
(b) Stroboscopic view of the motion. Each point corresponds to the
position and velocity measured at time intervals taken with a strobe 
frequency $\omega= (E_{3} - E_{1})/{\hbar}$.
(c) Power spectrum obtained from the time series of $x(t)$ as a
function of the frequency $f={\omega}/{2\pi}$.
}                                            
\label{fig:fig3}                                 
\end{center}                                 
\end{figure}

The time-dependent solution to the Schr\"odinger equation can be expressed by
a linear combination of the above eigenfunctions, namely,

\begin{equation}
\psi (x,t)= \sum_{n}c_{n}u_{n}(x)\exp(-E_{n}t/\hbar),
\label{eq:biwave}
\end{equation}
where $c_{n}$ are arbitrary complex coefficients. The motion of the 
particle will be dictated in part by the bistable potential and also
by the form of the initial wave packet.
We shall take  a linear
combination of the first few eigenfunctions as the initial wave packet. 
The system of units used is such that ${\hbar}=M=1$, and the lengths are
measured in units of $\beta^{-1}$ (i. e. $\beta= 1$).
For a given initial condition, the trajectory of the particle is obtained
by integrating the guidance formula, Eq. (\ref{eq:vel}).
The integration procedure is performed using a fourth-order Runge-Kutta
algorithm with a step $\Delta t= 0.001 $. 

\section{\label{sec:3.} Results and Conclusions}
Let us first consider the case in which the
initial wave packet is a linear combination of the ground state and the
third exited state,  
that is, ${\psi}(x,0)= iu_{0}(x) + 10\,u_{3}(x)$.  This particular choice 
leads to a quantum potential which for $t = 0$ is 
shown by the dotted line in Fig.~\ref{fig:fig1}(a).  
The bistable potential is the dashed line in the same figure.
As a result, the effective 
potential, shown in Fig.~\ref{fig:fig1}(a) by the solid line, has the
shape of a
triple-well with minima at the origin and near the two minima of
the classical potential  
($x=\pm [\ln(2 + \sqrt{3})]/2$).
An additional effect of the quantum 
potential is to flatten out the bistable potential in regions away from the
origin. Notice that the effective potential is time-dependent and its potential
wells will oscillate in time as dictated by the time dependent
wave packet. 
Under the action of the effective potential and for the initial 
condition $x= 0$, the particle will oscillate about the origin with frequency
$\omega= (E_{3} - E_{0})/{\hbar}$. From the classical point of view the 
motion is taking place about the {\it maximum} of the bistable potential.
We also observe periodic motions with the same frequency for initial 
positions near the points of minima of the classical potential.
Figure~\ref{fig:fig1}(b)
displays phase-space plots for initial particle's positions at left, center, 
and right of the potential.

A second choice for the initial wave packet,
${\psi}(x,0)= iu_{0}(x) + c_{1}u_{1}(x) + 10\,u_{3}(x) $,
with $c_{1}= 1$, leads to the motion
depicted in Fig.~\ref{fig:fig2}(a). The initial position is taken to be at $x= 0$.
The quantum potential is quite distinct from the previous one.
Its resulting effect
is to cause the motion to be more involved than in the preceding
case. The particle now moves from one side of the bistable potential to the
other, oscillating about the classical turning points
(around $x= \pm 0.5$) before moving
to the other side of the potential.  A stroboscopic plot of the motion is
shown in Fig.~\ref{fig:fig2}(b). Each point in the diagram corresponds to the position and
velocity measured with a strobe frequency of
$\omega= (E_{3}-E_{0})/{\hbar}$. The points lay in a single curve which
indicates that the motion is quasiperiodic. This is also verified by the 
Fourier analysis of the time-series for $x(t)$. The power spectrum,
Fig.~\ref{fig:fig2}(c), shows a sharply defined set of peaks.
The largest peak corresponds to
oscillations between the two lowest eigenstates, with frequency
$f_{10}= {\omega}_{10}/{2\pi}= (E_{1} - E_{0})/{2\pi\hbar}$.
The transition from quasiperiodic to periodic behavior is observed to
occur in a continuos way. 
As we reduce the value of the parameter $c_{1}$ towards zero, the Poincar{\'e} 
plot (Fig.~\ref{fig:fig2}(b)) gets smaller and eventually shrinks to a single point,
corresponding to the periodic motion of the central curve in Fig.~\ref{fig:fig1}(b).

\begin{figure}
\begin{center}
\includegraphics[height=5cm]{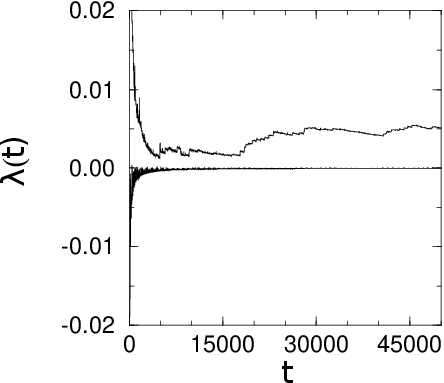}       
\caption
{
Upper curve: Largest Lyapunov exponent for a particle
in a bistable potential.
The length unit is taken to be $\beta^{-1}$, and in the system
of units used $\hbar=M=1$.
The time dependence of the mean 
rate of exponential separation $\lambda(t)$ was calculated from
Eq. (\ref{eq:exponent}) (with $t= n\Delta t$) along a trajectory 
generated by the same initial conditions as those of Fig.~\ref{fig:fig3}(a). 
Lower curve: Largest Lyapunov exponent for the quasi-periodic
 motion shown in Fig.~\ref{fig:fig2}.
}                                            
\label{fig:fig4}                                  
\end{center}                                 
\end{figure}

Finally, by considering the initial wave packet
${\psi}(x,0)= iu_{0}(x) + u_{1}(x) + c_{2}u_{2}(x) + 10\,u_{3}(x)$ with $c_{2}=4$,
and initial position $x= 0$, we obtain the trajectory depicted in Fig.~\ref{fig:fig3}(a). The 
motion is similar to that of the quasiperiodic case in the sense that it
takes place between both minima of the bistable potential, with oscillations
around one of the minima before returning to the other. However, this occurs 
in a rather disordered fashion. A stroboscopic view of the trajectory using
the same strobe frequency as in the quasiperiodic case, Fig.~\ref{fig:fig3}(b), reveals that the 
points are now scattered over a region of the phase-space. Also, the power
spectrum obtained from the Fourier analysis of the time series for $x(t)$, which
is depicted in Fig.~\ref{fig:fig3}(c), shows
a broad band spectrum distribution of frequencies in the background of a few sharp
peaks.  The plots in Figs.~\ref{fig:fig3}(b) and \ref{fig:fig3}(c)   indicate that the particle is undergoing 
a chaotic motion. 
The two largest peaks observed in the power spectrum correspond to the two
lowest frequencies of the system, namely 
$f_{10}= (E_{1} - E_{0})/{2\pi\hbar}$ and 
$f_{21}= (E_{2} - E_{1})/{2\pi\hbar}$. By reducing the parameter $c_{2}$, the 
points in the Poincar{\'e} plot become less and less scattered and
eventually they coalesce 
into the curve shown in Fig.~\ref{fig:fig2}(b). This shows that a transition from chaotic
to quasiperiodic behavior occurs in a continuos fashion as we vary
the initial wave packet in a suitable way. 
The sensitivity to initial conditions can be measured by considering two
neighboring trajectories separated, at a time $t_{i}= i{\Delta t}$, by a
distance $d_{i}$. Under the action of the flow, after a time interval
$\Delta t$, 
the separation of the two trajectories becomes $d_{i+1}$. 
The mean rate of exponential separation between two neighboring trajectories is 
given by:

\begin{equation}
\lambda(n)= {1\over{n\Delta t}}\sum_{i=1}^{n}\ln({d_{i}\over d_{i-1}})
\label{eq:exponent}
\end{equation}
where $n\Delta t$ is the total integration time. The largest Lyapunov 
exponent is determined by the asymptotic value of the mean rate of 
exponential separation, namely $\lambda= \lim_{n\to\infty} \lambda(n)$.
By taking $d_{i}= d_{0}= 10^{-6}$ and ${\Delta t}= 0.001$, we obtain 
$\lambda\simeq 0.005$ for $n$ as large as $5{\times}10^7$. The time
dependence of Eq. (\ref{eq:exponent}) is shown in Fig.~\ref{fig:fig4}. The positive 
value for the Lyapunov exponent is an indication that the motion is 
indeed chaotic. For comparison we have also plotted the largest Lyapunov
exponent for the motion depicted in Fig.~\ref{fig:fig2}. The exponent is clearly zero,  
as we expected in the case of quasi-periodic motion.

In conclusion, using Bohm's quantum mechanics, we have studied the dynamics of
a particle in the presence of a one-dimensional bistable potential.  Contrary
to what have been previously suggested, we found that the particle may exhibit
chaotic behavior, depending on the initial form of the wave function 
as well as on its initial position. 
A judicious choice of initial wave functions allowed us to
investigate the transition from quasiperiodic to chaotic motion, 
as well as from
quasiperiodic to periodic motion. 
In both cases the transition between different 
regimes is accomplished in a continuos manner, as inferred from the 
Poincar{\'e} plots and Fourier spectral analysis of the motion.



\end{document}